# Quantum Coherence Engines

Elias P. Gyftopoulos
Ford Professor Emeritus of the Departments of
Mechanical and Nuclear Engineering
Massachusetts Institute of Technology

In a report published in Science, Scully et al propose to generate "phaseonium", and "extract work from a single heat bath via vanishing coherence" [1]. On the basis of the laws of physics, such a proposal is futile for at least two reasons.

(1) The correct terminology for a heat bath is simply a *reservoir* because any system in any state (thermodynamic equilibrium or not thermodynamic equilibrium) contains neither heat nor work. Heat and work are not properties of a system. They are modes of interactions. A reservoir is a system that experiences only heat interactions with other systems, that is, interactions that involve the exchange of both energy $Q_R$ (or $\delta Q_R$), and entropy $Q_R/T_R$ (or $\delta Q_R/T_R$), where $T_R$ is constant for all values of the energy $Q_R$, and $Q_R$ and $\delta Q_R$ are independent of $T_R$ [2]. It follows that no work can be extracted from a reservoir because work is an interaction that involves the exchange of energy $W$ (or $\delta W$) only. If work could be extracted from a reservoir, then Maxwell's demon would be feasible, a feasibility that for the first time has been recently shown to be impossible [3].

(2) The second reason is even more important, and is provided by Scully himself [4]. He estimates that in order to achieve work output $W_{net}$ from "phaseonium", he must use devices that consume more than 1,000 $W_{net}$ of energy. It follows that the gain $W_{net}$ from "phaseonium" is not due to extraction of work from a single bath but to the consumption of a very large amount of energy. It is noteworthy that, despite the complete lack of the relevant scientific principles, even the first engines that were built about two and a half centuries ago had a better efficiency [5] than the proposed scheme for the generation of "phaseonium".

References
[1] M. O. Scully, M. Suhail Zubairy, G. S. Agarwal, H. Walther, *Science*, 299, 862 (2003).
[2] E. P. Gyftopoulos, G. P. Beretta, *Thermodynamics: Foundations and Applications*, Macmillan, New York (1991).
[3] E. P. Gyftopoulos, *Physica A*, 307, 421 (2002).
[4] M. O. Scully, in *Quantum Limits to the Second Law*, D. P. Sheehan, Ed., American Institute of Physics, Melville, NY (2002).
[5] D. S. L. Cardwell, "From Watt to Clausius", Cornell University Press, Ithaca, NY (1971).